\documentclass{acta}

\usepackage{amsmath}
\usepackage{amssymb}
\usepackage{rotating}
\usepackage[hidelinks]{hyperref} 
\usepackage{bm}

\usepackage{xcolor}

\newcommand{\ra}[4]{${#1}^{\rm h}{#2}^{\rm m}{#3}\fs{#4}$}
\newcommand{\dec}[4]{${#1}\arcdeg{#2}\arcmin{#3}\farcs{#4}$}

\newcommand\fs{\mbox{$.\!\!^{\mathrm s}$}}%
\newcommand\arcdeg{\mbox{$^\circ$}}%
\newcommand\arcmin{\mbox{$^\prime$}}%
\newcommand\farcs{\mbox{$.\!\!^{\prime\prime}$}}%

\newcommand{\dummy}[1]{}

\begin{document}

\begin{Titlepage}

\Title{\textit{Eppur non si trovano Vol.~3}:\\
Phoebe  -- a Mirage of a Primordial Black Hole}

\Author{Andrzej Udalski~~ and~~ Przemek Mr\'oz}
{Astronomical Observatory, University of Warsaw, Al. Ujazdowskie 4, 00-478 Warszawa, Poland\\
e-mail: udalski@astrouw.edu.pl, pmroz@astrouw.edu.pl}

\Received{MM DD, YYYY}

\end{Titlepage}

\vspace*{-10pt}
\Abstract{
Recent preprints by Key {\it et al.} reported the discovery of a short-lived brightening of a star (nicknamed `Phoebe') located in the Large Magellanic Cloud that was interpreted as a short-timescale gravitational microlensing event produced by a lunar-mass primordial black hole (PBH) in the Milky Way dark matter halo. Here, we present an independent re-analysis of the publicly available DECam observations of this object, incorporating additional data from 2020 and 2021. The object underwent at least three distinct, low-amplitude brightenings (one of which was misinterpreted as a short-timescale microlensing event) in addition to long-term variations of its mean magnitude. These characteristics indicate that Phoebe is an ordinary variable star rather than a microlensing event. This finding resolves the apparent tension with the results from earlier microlensing experiments that rule out the hypothesis that a substantial fraction of dark matter is composed of lunar- and planetary-mass PBHs.
}
{Gravitational microlensing (672), Dark matter (353), Milky Way dark matter halo (1049), Primordial black holes (1292), Large Magellanic Cloud (903)}

\section{Introduction} \label{sec:intro}

Recent preprints by \citet{key2026a,key2026b} presented an analysis of high-cadence photometric observations of the Large Magellanic Cloud (LMC) that were collected by the Dark Energy Camera (DECam) over five consecutive nights in December 2019. Approximately 4~million stars were monitored at a sub-minute cadence of 50\,s. \citet{key2026a,key2026b} reported the discovery of a single short-lived brightening in the light curve of a faint ($g=21.547 \pm 0.010$, $r=21.285 \pm 0.012$; magnitudes are not corrected for extinction; \citealt{nidever2021}) star, nicknamed Phoebe. Its J2000 equatorial coordinates are (R.A., Decl.) = (\ra{05}{52}{25}{67}, \dec{-70}{44}{21}{9}). \citet{key2026a,key2026b} interpreted this signal as a short-timescale gravitational microlensing event with an estimated Einstein timescale of $0.043^{+0.010}_{-0.007}$\,d, making it one of the shortest on record. The authors considered various possible locations of the purported lens: in the Milky Way disk, the LMC disk, and the Milky Way or LMC dark matter haloes. Based on a Bayesian analysis, \citet{key2026a,key2026b} concluded that the dark matter halo interpretation was correct, claiming that the short-lived signal represents the discovery of a lunar-mass primordial black hole (PBH) in the Milky Way dark matter halo. They argued that this finding indicates that dark matter is composed of compact, lunar-mass objects, possibly PBHs created during the epoch of cosmological inflation.

This claim stands in stark tension with the results of the previous high-cadence microlensing experiments targeting the Magellanic Clouds. In particular, the Optical Gravitational Lensing Experiment (OGLE; \citealt{udalski2015}) has monitored a 12.6 deg$^2$ area of the Magellanic Clouds at a high cadence (16 to 20\,min) since 2022. The results of the searches for short-timescale microlensing events in the first two years of these observations were published by \citet{mroz2024d}. A total of about 35~million source stars were observed at high cadence over approximately 435 nights. No convincing short-timescale microlensing events were found in the OGLE data, placing strong limits on the abundance of lunar- and planetary-mass PBHs in the Milky Way dark matter halo. In particular, if the entirety of dark matter were composed of lunar-mass ($10^{-7}$\,M$_\odot$) PBHs, OGLE should have detected approximately 1500 events. 

Conversely, if the candidate identified by \citet{key2026a,key2026b} were a genuine microlensing event produced by a lunar-mass PBH, by scaling by the number of monitored source stars and total light curve duration, OGLE should have detected $\approx 700$ similar events, assuming comparable detection efficiencies for Phoebe-like events between the OGLE and DECam surveys. 

In addition, the light curve of the purported microlensing event presented in Figure~1 of \citet{key2026b} shows some suspicious features. For example, on the night of 2019 Dec 16/17, the star appears to be magnified by $\approx 0.05$\,mag relative to its baseline on other nights. This systematic departure from a constant baseline indicates that the source may be intrinsically variable, casting doubt on the reality of the microlensing event.

Given the profound implications of confirming a short-timescale microlensing event toward the Magellanic Clouds for fundamental physics and astronomy, we decided to re-analyze the publicly available DECam data with an independent photometric pipeline. This letter presents our findings.

\begin{table}
\caption{Log of DECam observations}
\label{tab:log}
\centering
\begin{tabular}{lrrlrr}
\hline \hline
\multicolumn{1}{c}{Night} & \multicolumn{1}{c}{Images} & \multicolumn{1}{c}{FWHM} & \multicolumn{1}{c}{Night} & \multicolumn{1}{c}{Images} & \multicolumn{1}{c}{FWHM}\\
\hline
2019-12-15/16 & 569 & 1\zdot\arcs41 & 2020-02-21/22 & 270 & 0\zdot\arcs92 \\ 
2019-12-16/17 & 540 & 1\zdot\arcs05 & 2020-02-22/23 & 353 & 1\zdot\arcs03 \\
2019-12-17/18 & 561 & 0\zdot\arcs96 & 2021-12-10/11 &  19 & 1\zdot\arcs23 \\
2019-12-18/19 & 540 & 0\zdot\arcs96 & 2021-12-11/12 &  17 & 1\zdot\arcs46 \\
2019-12-19/20 & 423 & 1\zdot\arcs23 & 2021-12-12/13 &   8 & 1\zdot\arcs61 \\
2020-02-13/14 & 227 & 1\zdot\arcs10 & 2021-12-13/14 &  15 & 1\zdot\arcs74 \\
2020-02-14/15 & 166 & 1\zdot\arcs06 & 2021-12-14/15 &  18 & 1\zdot\arcs40 \\
2020-02-16/17 &  48 & 1\zdot\arcs21 & 2021-12-26/27 &   4 & 1\zdot\arcs28 \\
2020-02-17/18 & 349 & 1\zdot\arcs10 & 2021-12-27/28 &  11 & 1\zdot\arcs18 \\
2020-02-18/19 &   1 & 1\zdot\arcs36 & 2021-12-28/29 &  15 & 1\zdot\arcs04 \\
2020-02-19/20 & 353 & 1\zdot\arcs24 & 2021-12-29/30 &  16 & 1\zdot\arcs40 \\
2020-02-20/21 & 273 & 0\zdot\arcs93 & 2021-12-30/31 &   8 & 1\zdot\arcs22 \\
\hline
\end{tabular}
\end{table}

\section{Data}


\citet{key2026a,key2026b} used images collected by DECam \citep{flaugher2015} mounted on the Victor M.~Blanco 4-m Telescope at the Cerro Tololo Inter-American Observatory, Chile. The camera consists of 62 science CCD detectors. Each detector has $2046 \times 4094$ pixels, providing a mean plate scale of $0.27''$ per pixel and a total field of view of 3\,deg$^2$. The short-timescale microlensing candidate is located on the detector \#51 (N20). 

We downloaded the publicly available DECam images from programs 2019B-0071, 2020A-0913, and 2021B-0023 (PI: J.~Mould) from the NOIRLab Astro Data Archive\footnote{https://astroarchive.noirlab.edu/}. \citet{key2026a,key2026b} restricted their analysis to data from program 2019B-0071, which spanned five nights from 2019 Dec 15/16 through Dec 19/20. The 2020A-0913 and 2021B-0023 programs totaled 5 and 8.5 nights, respectively. These additional images were collected from 2020 Feb 13/14 to Feb 22/23 and 2021 Dec 10/11 to 30/31. Some images taken from program 2020A-0913 and most frames from program 2021B-0023 appear to have been acquired with the telescope tracking along the declination axis. Given the large number of trailed images, we suspect that this setup was intentional, likely aiming to search for possible gravitational microlensing events produced by asteroid-mass PBHs whose expected Einstein timescales are shorter than a typical exposure time (20 s). However, if true, such an observing strategy was fundamentally limited by finite-source effects: for events produced by low-mass lenses ($\lesssim 10^{-6}$\,M$_\odot$) the event duration is predominantly set by the crossing timescale of the source star's disk rather than the Einstein timescale, the amplitude of the event falls below the detection limit, not to mention wave-optics effects (the Schwarzschild radius of an asteroid-mass PBH is smaller than the wavelength of visible light). Because our photometric pipeline was not designed to process trailed stellar images, we excluded them from the subsequent analysis.

All analyzed images were taken with 20\,s exposures in the broadband filter $VR$, whose transmission curve covers the wavelength range 500--760\,nm. The raw data were processed using the DECam Community Pipeline \citep{valdes2014}, which includes bias, dark, linearity, and flat-field corrections, as well as astrometric calibration. The full log of the DECam observations analyzed in this work is presented in Table~\ref{tab:log}.

\begin{figure}
\centering
\includegraphics[width=\textwidth]{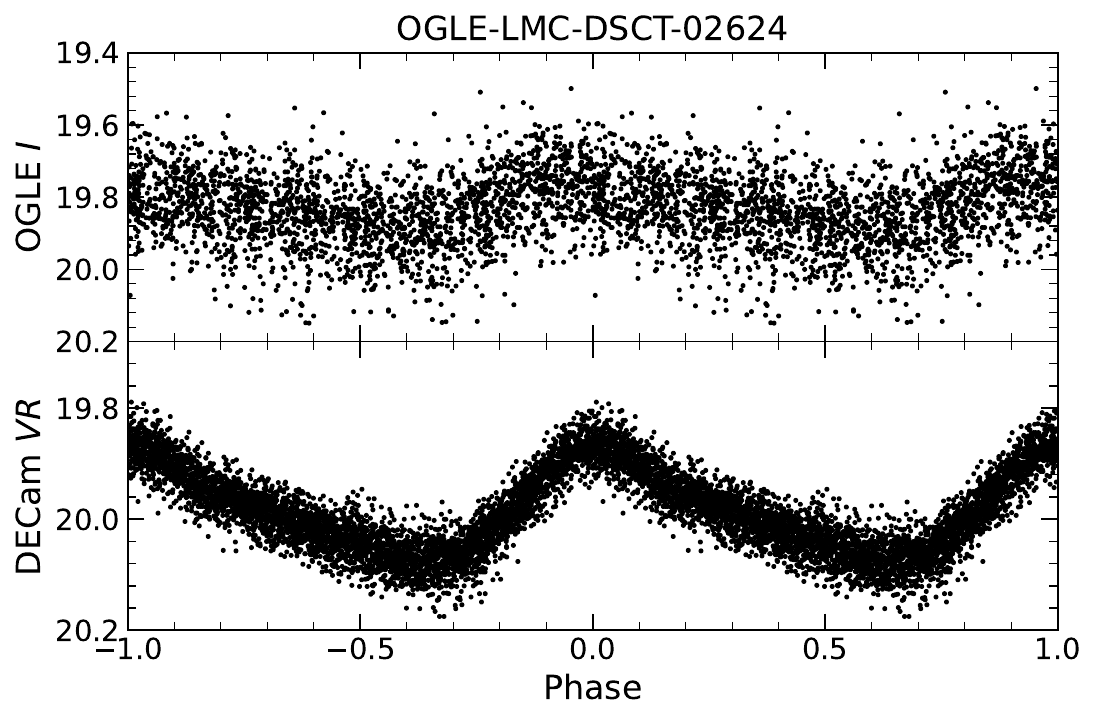}
\caption{Comparison between the OGLE (upper panel) and DECam (lower panel) light curves of a $\delta$ Scuti star OGLE-LMC-DSCT-02624. The pulsation period is 0.11883125(5)\,d. The upper and lower panels have $y$-axis ranges of 0.8 and 0.5\,mag, respectively.}
\label{fig:dsct}
\end{figure}

\section{DIA Photometry}

The CCD detectors of the DECam mosaic camera closely resemble those of the \mbox{OGLE-IV} camera \citep{udalski2015}. Their physical dimensions are nearly identical, namely $2046 \times 4094$ pixels for DECam \vs $2048\times 4102$ pixels for OGLE. The pixel scale on the sky is also remarkably similar: 0.27 \vs 0.26 arcsec\,pixel$^{-1}$ for DECam and OGLE cameras, respectively. It was therefore a natural choice to reduce the DECam images of the dense stellar LMC fields using the well-tested OGLE photometric pipeline \citep{udalski2003}, which is based on the Difference Image Analysis (DIA; \citealt{alard1998}) technique implemented by \citet{wozniak2000}. This pipeline has been used routinely by the OGLE survey since the start of its third phase in 2001 and is widely recognized for providing the most accurate ground-based survey photometry. Adapting the OGLE pipeline for DECam images required only minor fine-tuning of the software parameters.

In general, standard DIA reduction steps, such as those described by \citet{udalski2015} and \citet{mroz2026}, were applied to the individual DECam images. A deep, nearly noiseless reference image was constructed by co-adding 20 individual frames with the highest signal-to-noise ratio and best resolution (seeing), collected during the nights of 2019 Dec 17/18 and 18/19. This reference image was subsequently used for the reductions of all DECam frames.

Some of the DECam images were taken under poor atmospheric conditions, \textit{i.e.}, through clouds (sometimes quite thick) or during periods of poor seeing. To avoid low-quality measurements, we applied a quality cut based on a simple filter measuring the signal level of bright, constant stars in the field. Frames in which the mean signal dropped by more than 0.75\,mag (corresponding to a factor of two) relative to optimal transparency conditions were discarded.

As the DECam CCD~\#51 (N20), where Phoebe is located, covers a relatively crowded stellar region of the LMC (corresponding to the OGLE field LMC552.09), we decided to use similar subframing during the reductions as during the OGLE reductions of this area, namely the chip was divided into a $4\times 2$ grid of $1{\rm k}\times 1{\rm k}$ subframes. Each subframe was reduced individually, and the resulting photometry from subframes was then merged into one final photometry file for a given frame. This file contained the brightness measurements for all stars detected on the reference image. Finally, once the photometric reductions of all DECam images were complete, a standard OGLE-style database \citep{udalski2015} was built for easy data access and analysis. One of our quality tests -- a comparison between the OGLE and DECam light curves of a $\delta$ Scuti star OGLE-LMC-DSCT-02624 \citep{soszynski2023}, located near Phoebe in the sky, is presented in Figure~\ref{fig:dsct}.

Because the DECam images were collected through a wide $VR$ filter, it was not possible to precisely calibrate the photometry. Instead, we decided to add a constant zero-point shift to our instrumental magnitudes to approximate the $R$-band magnitudes. We derived this shift by comparing the mean instrumental magnitudes of a few stars with their known standard $V$- and $I$-band photometry from the OGLE catalog.

\begin{figure}
\centering
\includegraphics[width=0.5\textwidth]{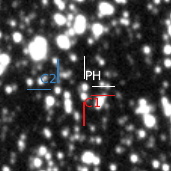}
\caption{$30'' \times 30''$ cutout of the reference image around Phoebe (PH). The positions of Phoebe and the two comparison stars (C1, C2) are indicated by tick marks. North is up, and East is to the left.}
\label{fig:fc}
\end{figure}

\begin{figure}
\includegraphics[width=\textwidth]{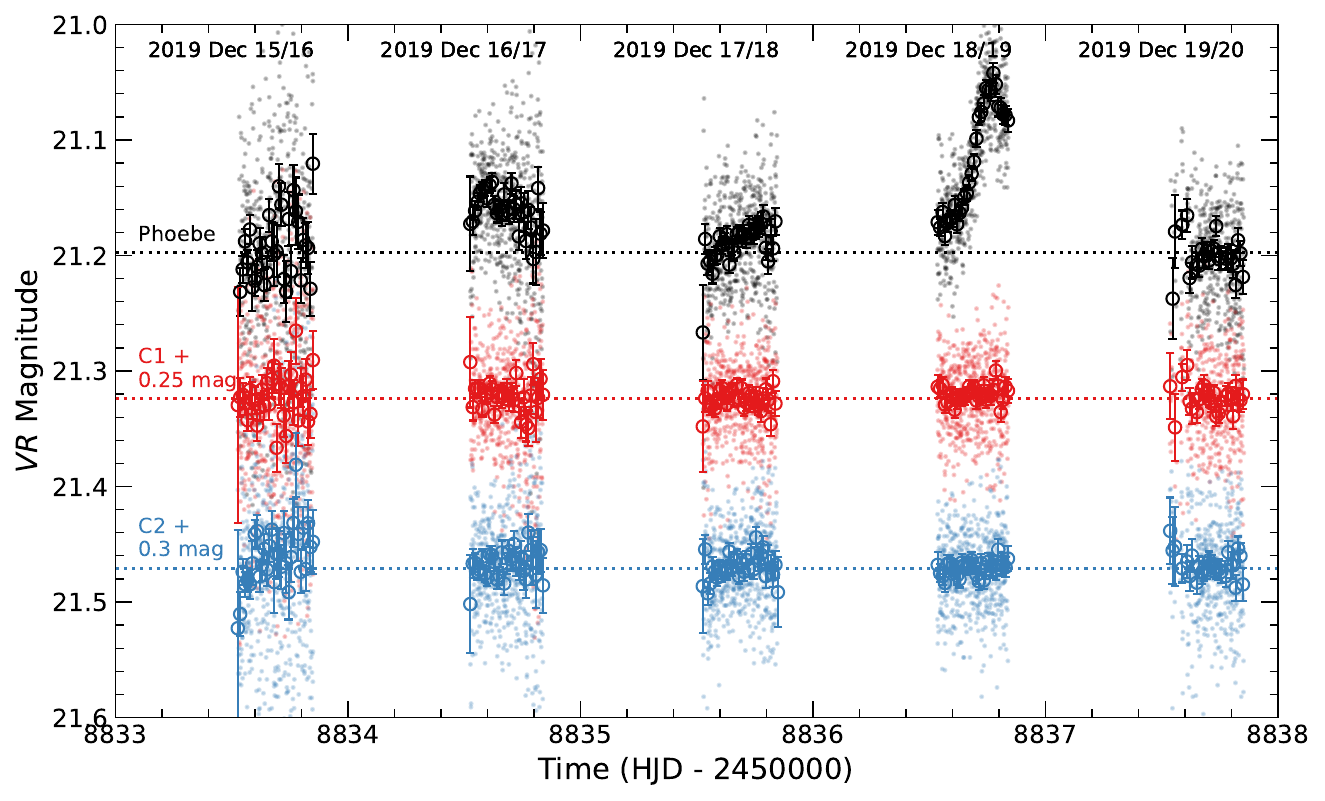}
\caption{Light curves of Phoebe and the two comparison stars from program 2019B-0071.}
\label{fig:lc1}
\end{figure}

\begin{figure}
\includegraphics[width=\textwidth]{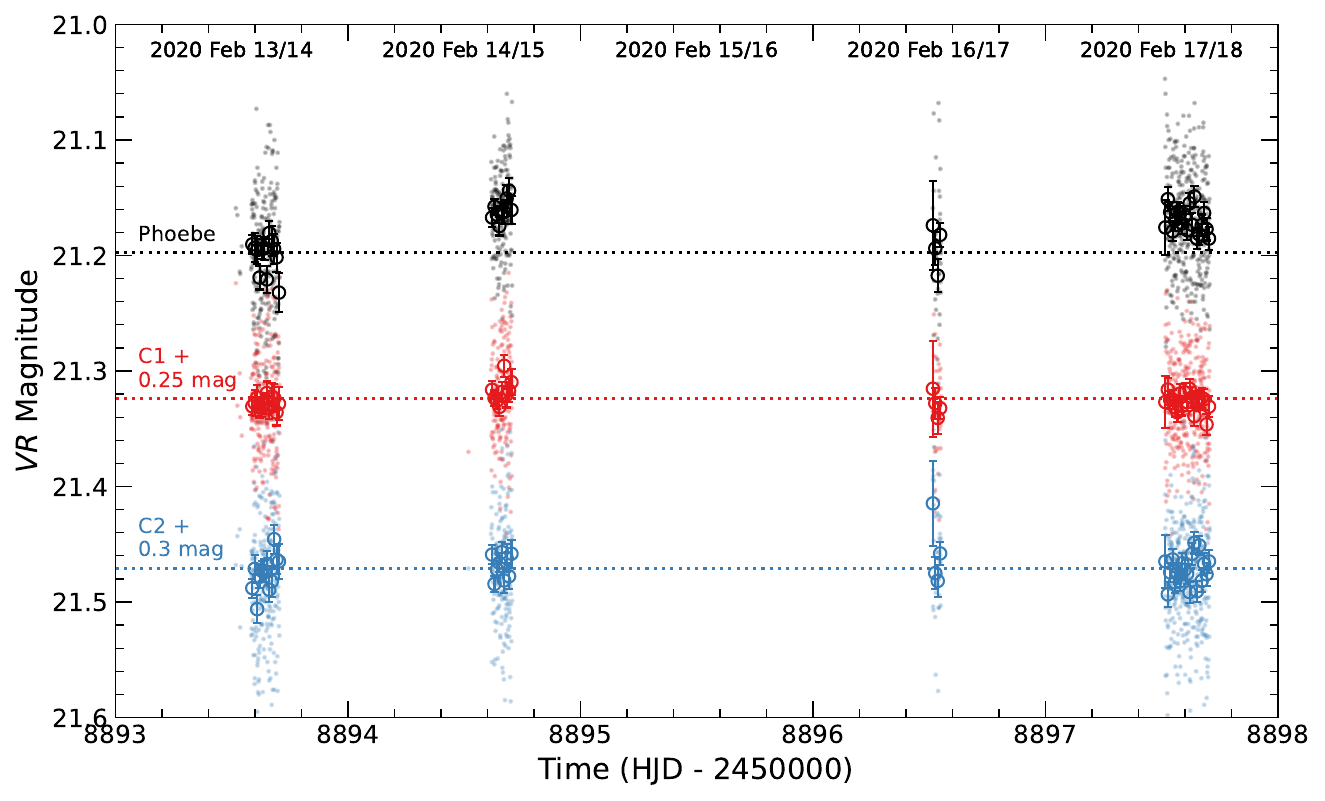}
\includegraphics[width=\textwidth]{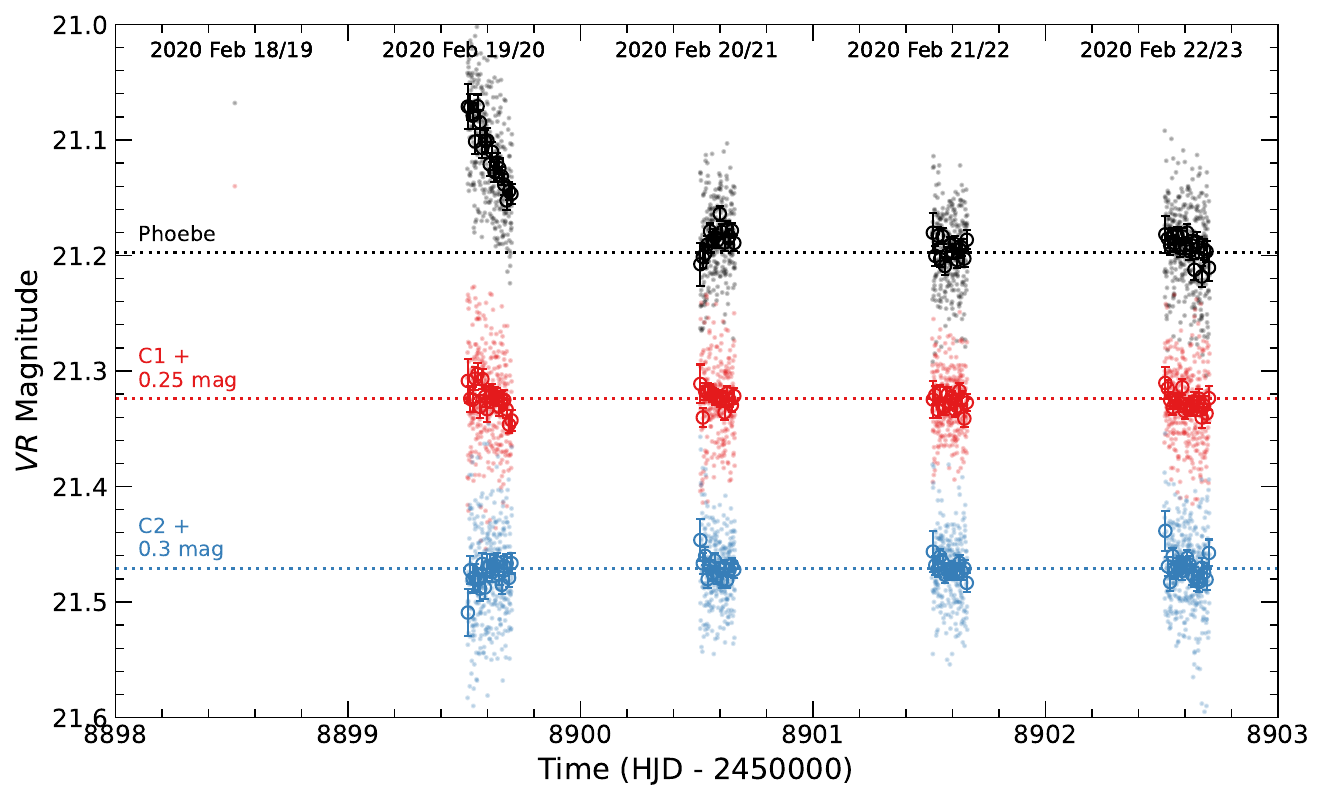}
\caption{Light curves of Phoebe and the two comparison stars from program 2020A-0913.}
\label{fig:lc2}
\end{figure}

\begin{figure}
\includegraphics[width=\textwidth]{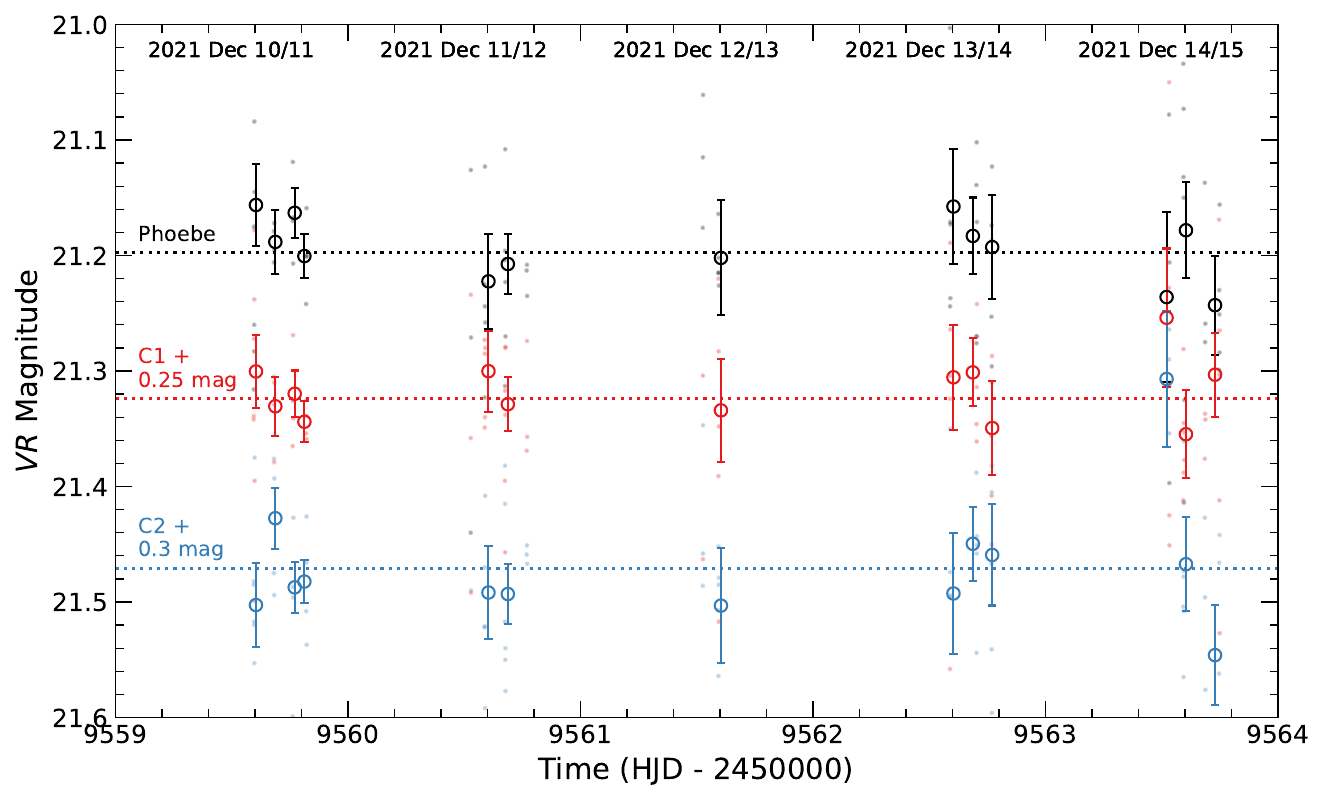}
\includegraphics[width=\textwidth]{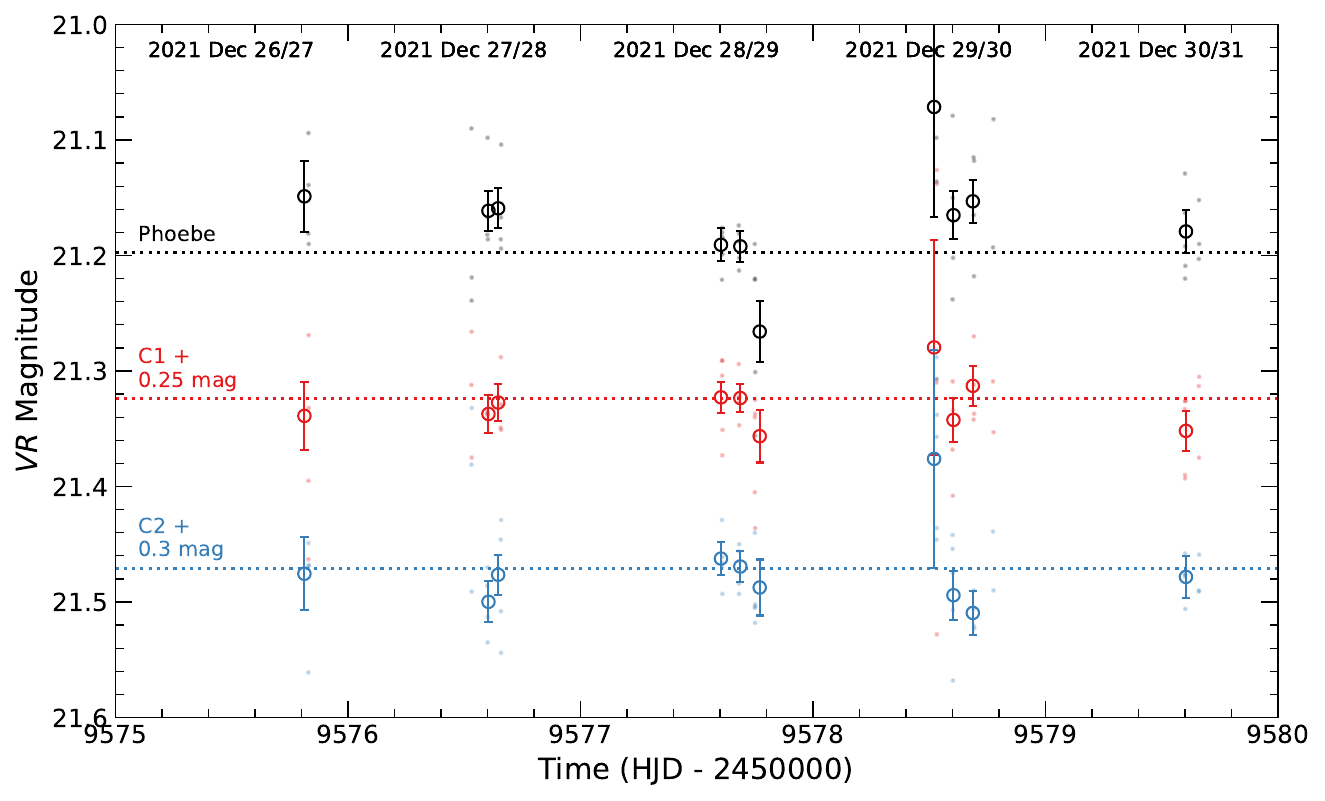}
\caption{Light curves of Phoebe and the two comparison stars from program 2021B-0023.}
\label{fig:lc3}
\end{figure}

\section{Results and Conclusions}

Figure~\ref{fig:fc} shows a $30'' \times 30''$ cutout of the reference image centered on the target (Phoebe). The positions of Phoebe (PH) and two constant comparison stars of similar brightness (C1 and C2) are indicated by tick marks.  

Figures~\ref{fig:lc1}--\ref{fig:lc3} present the DIA light curves of Phoebe and the comparison stars. Individual measurements are plotted as small dots. Because none of the three objects exhibit significant variability on timescales shorter than a few minutes, the data were binned using 15-minute intervals to improve the signal-to-noise ratio (60-minute intervals for the 2021 data). The black dotted line indicates the mean magnitude of Phoebe, calculated using data from the nights of 2019 Dec 15/16 and 19/20. The red and blue dotted lines denote the mean magnitudes of the comparison stars (corresponding to C1 and C2, respectively) that were calculated using all available data.

Figure~\ref{fig:lc1} shows the data from program 2019B-0071, i.e., the same as those analyzed by \citet{key2026a,key2026b}. We were able to successfully recover the brightening detected on the night of 2019 Dec 18/19. Unfortunately, the falling branch of the light curve is missing, rendering it impossible to evaluate the symmetry of the signal, which is a crucial diagnostic for its classification as a microlensing event. In addition, Phoebe also exhibited significant variations on other nights, Dec 16/17 and possibly Dec 17/18. Notably, the variability on the night of Dec 16/17 is clearly visible in the light curves presented by \citet{key2026a,key2026b}.

Figures~\ref{fig:lc2} and \ref{fig:lc3} show the extended light curves from programs 2020A-0913 and 2021B-0023. The star underwent another brightening on the night of 2020 Feb 19/20. Furthermore, its mean brightness changed across the full duration of the DECam monitoring campaign. 

Our full light curve of Phoebe is publicly available \textit{via}
\begin{center}
\textit{https://zenodo.org/records/20735480}
\end{center}

One may argue that the second brightening seen in the 2020 data is another short-timescale microlensing event produced by another asteroid-mass PBH. The two brightenings are separated by approximately 63 days, implying the event rate higher than 10 per year per star if both were microlensing events. Therefore, considering that the 2019 program observed approximately $4 \times 10^6$ stars for effectively 1.6 days, about $10 \times 4 \times 10^6 \times 1.6 / 365.25 \approx 180,000$ Phoebe-like events would have been seen in light curves of other stars, which is definitely excluded by the data.

Our independent re-analysis of the DECam data demonstrates that Phoebe is variable on all timescales. It underwent at least three distinct, low-amplitude brightenings (one of which was misinterpreted by \citet{key2026a,key2026b} as a short-timescale microlensing event) in addition to long-term variations of the mean brightness. These observational characteristics indicate that Phoebe is an ordinary variable star, rendering the claims of the discovery of a lunar-mass PBH in the Milky Way dark matter halo invalid.

This is not the first time that a variable star has been mistaken for a short-timescale microlensing event in high-cadence time-series observations of limited duration \citep{mroz2023b,mroz2026}. Extended, long-timescale (months to years) monitoring is needed to robustly distinguish true gravitational microlensing events from intrinsic stellar variability in the photometric data.

\section*{Acknowledgments}

We thank members of the OGLE team for discussions about the DECam microlensing event candidate and Raphael Oliveira for his help in accessing the SMASH survey data. This research was funded in part by National Science Centre, Poland, grant SONATA 2023/51/D/ST9/00187 awarded to P.M. A.U. acknowledges support from the grant OPUS-28 2024/55/B/ST9/00447 of the Polish National Science Centre. This research uses services or data provided by the Astro Data Archive at NSF's NOIRLab. NOIRLab is operated by the Association of Universities for Research in Astronomy (AURA), Inc.~under a cooperative agreement with the National Science Foundation. For the purpose of Open Access, the authors have applied a CC-BY public copyright license to any Author Accepted Manuscript (AAM) version arising from this submission.

\bibliographystyle{acta}
\bibliography{pap}

\end{document}